\documentclass[12pt]{article}
\usepackage{graphicx}
\usepackage{amssymb}
\usepackage{amsmath}

\newcommand{\bear}{\begin{array}}  
\newcommand {\eear}{\end{array}}
\newcommand{\bea}{\begin{eqnarray}}   
\newcommand{\eea}{\end{eqnarray}}
\newcommand{\beq}{\begin{equation}}   
\newcommand{\eeq}{\end{equation}}
\newcommand{\bef}{\begin{figure}}  \newcommand 
{\eef}{\end{figure}}
\newcommand{\bec}{\begin{center}}  \newcommand 
{\eec}{\end{center}}

\def\lrfp#1#2#3{ \left(\frac{#1}{#2} 
\right)^{#3}}

\def\lrfp#1#2#3{ \left(\frac{#1}{#2} 
\right)^{#3}}

\begin{document}

\begin{titlepage}

\begin{flushright}
IPMU 08-0069 \\
ICRR-Report-529
\end{flushright}

\vskip 1.35cm

\begin{center}
{\large \bf
A General Analysis of Non-Gaussianity from \\Isocurvature Perturbations
}
\vskip 1.2cm

Masahiro Kawasaki$^{a,b}$,
Kazunori Nakayama$^a$,
Toyokazu Sekiguchi$^a$,
Teruaki Suyama$^a$ and
Fuminobu Takahashi$^b$

\vskip 0.4cm

{ \it $^a$Institute for Cosmic Ray Research,
University of Tokyo, Kashiwa 277-8582, Japan}\\
{\it $^b$Institute for the Physics and Mathematics of the Universe,
University of Tokyo, Kashiwa 277-8568, Japan}
\date{\today}

\begin{abstract}
Light scalars may be ubiquitous in nature, and their quantum fluctuations
 can produce large non-Gaussianity in the cosmic
microwave background temperature anisotropy. The non-Gaussianity may be
accompanied with a small admixture of isocurvature perturbations, which
often have correlations with the curvature perturbations.
We present a general method to calculate the non-Gaussianity
in the adiabatic and  isocurvature perturbations with and without correlations, and see how it works
in several explicit examples. We also show that they leave distinct signatures on the bispectrum of the cosmic microwave background temperature fluctuations.
\end{abstract}


\end{center}
\end{titlepage}

\section{Introduction} \label{intro}

The inflationary paradigm has received strong observational support
especially from the WMAP observation of the cosmic microwave
background (CMB) \cite{Komatsu:2008hk}; in a simple class of the
single-field inflation models, density fluctuations produced by an
inflaton are known to be nearly scale-invariant, adiabatic and
Gaussian, and these properties are found to be in a good agreement with the
observation.

In the standard picture, the inflationary expansion is driven by a
scalar field $\phi$, the inflaton, which slow-rolls on a very flat
scalar potential.  During inflation the inflaton $\phi$ acquires
quantum fluctuations, which result in slight differences in the
subsequent evolution at different places in the Universe. After
inflation those differences turn into spatial inhomogeneities in the
energy density.  The nearly scale-invariant, adiabatic and Gaussian
density perturbations are often regarded as the standard lore of the
inflation theory. However, we would like to emphasize  that the lore is based
on a simple but crude assumption that it is only the inflaton that
acquires sizable quantum fluctuations during inflation.  Its apparent
success does not necessarily mean that such a non-trivial
condition is commonly met in the landscape of the inflation theory.

 In fact, there are many flat directions in a supersymmetric (SUSY)
 theory and the string theory, and it may be even natural to expect
 that some of them remain light during inflation. If this is the case,
 such light scalars acquire quantum fluctuations, which may leave
 their traces in the CMB anisotropy such as isocurvature
 perturbations. This is indeed the case if those light
 scalars participate in the production of dark matter and/or baryon
 asymmetry of the Universe. For instance, the QCD axion \cite{Peccei:1977hh}, 
 which was proposed to solve the strong CP problem, is a plausible candidate for
 the cold dark matter (CDM), and it is known that the
 axion generally acquires quantum fluctuations leading to the the CDM
 isocurvature perturbations.  Also there are baryogenesis scenarios
 that contain light scalars. In the Affleck-Dine (AD) mechanism \cite{Affleck:1984fy}, for
 instance, a phase component of the AD field remains flat in most
 inflation models, leading to the baryonic isocurvature
 perturbations \cite{Linde:1985gh,Enqvist:1998pf,Kasuya:2008xp,Riotto:2008gs}.
  Although the observed density perturbation is almost
 adiabatic and no sizable isocurvature perturbation has been
 discovered so far, a small admixture of the isocurvature
 perturbations is still allowed.

Non-Gaussianities have recently attracted much attention since Yadav
and Wandelt claimed an evidence of the significant non-Gaussianity in
the CMB anisotropy data \cite{Yadav:2007yy}. On the other hand, the
latest WMAP five-year result was shown to be consistent with the
vanishing non-Gaussianity \cite{Komatsu:2008hk}, although the
likelihood distribution of the WMAP result seems to favor some amount
of non-Gaussianity.  Also there are active studies searching for the
non-Gaussianity \cite{Slosar:2008hx}, and it is not settled yet
whether the non-Gaussianity exists or not.  At the present stage, it
is fair to say that the observations are consistent with the nearly
scale invariant and pure adiabatic perturbations with Gaussian
statistics, while there is a hint of non-Gaussianity at the two sigma
level.

Suppose that there is indeed significant amount of non-Gaussianity.
Since it is known that  the slow-roll inflation with a canonical kinetic term generally predicts a
negligible amount of non-Gaussianity \cite{Acquaviva:2002ud,Maldacena:2002vr,Seery:2005wm,Yokoyama:2007uu},
we need to go beyond the simplest class of inflation models. A simple and even plausible way  is
to introduce additional light scalars. In
the curvaton \cite{Mollerach:1989hu,Linde:1996gt,Lyth:2001nq} and/or
ungaussiton \cite{Suyama:2008nt} scenarios, those light
scalars can generate sizable non-Gaussianity \cite{Lyth:2002my,Lyth:2005fi,Lyth:2006gd}. 
In the presence of additional light scalars
with quantum fluctuations, it is generically expected that isocurvature perturbation may arise.
In particular, the non-Gaussianity hinted by the recent observations may originate from
a small admixture of isocurvature perturbations \cite{Linde:1996gt,Bartolo:2001cw,Boubekeur:2005fj}. 

We recently presented a formulation on non-Gaussianity in the
isocurvature perturbations, and studied in detail how it exhibits
itself in the CMB temperature anisotropy \cite{Kawasaki:2008sn}. 
In Ref.~\cite{Kawasaki:2008sn}, we found that the non-Gaussianity in the
isocurvature perturbations leave distinctive signatures in the CMB; 
the non-Gaussianity is enhanced at large scales.
Such features will enable us to distinguish the non-Gaussianity in isocurvature perturbations
from that mainly in the adiabatic perturbation.
As an example we considered the non-Gaussianity in the CDM
isocurvature perturbations \cite{Kawasaki:2008sn} and the baryonic isocurvature
perturbations \cite{Kawasaki:2008jy}.

In this paper we extend our previous study in order to include 
possible correlations between the curvature and the isocurvature
perturbations. We will give explicit examples in which there actually
exist such correlations. Furthermore, as we did in the previous paper,
we will present how the CMB temperature anisotropies are affected by
the presence of the non-Gaussianities in the isocurvature
perturbations correlated with the curvature perturbations.

This paper is organized as follows. In Sec.~\ref{sec:formalism} 
we extend our formalism to include correlation of adiabatic and isocurvature perturbations.
In Sec.~\ref{sec:app} this formalism is applied to some explicit models.
We study features in the bispectrum of CMB anisotropy in Sec.~\ref{sec:CMB}.
Sec.~\ref{sec:conclusion} is devoted to discussion and conclusions.


\section{Formalism} \label{sec:formalism}

In this section,
we extend the formalism developed in \cite{Kawasaki:2008sn},
where the formulation to calculate the non-Gaussianity of the isocurvature
perturbation was provided,
to include more general case that isocurvature perturbations
have correlations with adiabatic one.
To be definite,
we consider CDM isocurvature perturbation, but
an extension to other types of the isocurvature perturbations is straightforward.

\subsection{Non-linear isocurvature perturbations and constraints}

We write the perturbed spacetime metric as
\begin{equation}
	ds^2=-{\mathcal N}^2 dt^2 +a^2(t)e^{2\psi} \gamma_{ij} \left ( dx^i + \beta^i dt \right )
	\left ( dx^j + \beta^j dt \right ),
\end{equation}
where ${\mathcal N}$ is the lapse function, $\beta_i$ the shift
vector, $\gamma_{ij}$ the spatial metric, $a(t)$ the background scale
factor, and $\psi$ the curvature perturbation.  On sufficiently large
spatial scales, the curvature perturbation $\psi$ on an arbitrary slicing at
$t=t_f$ is expressed by~\cite{Lyth:2004gb}
\begin{eqnarray}
	\psi (t_f,{\vec x})=N(t_f,t_i;{\vec x})-\log \frac{a(t_f)}{a(t_i)},
	 \label{curvature1}
\end{eqnarray}
where the initial slicing at $t=t_i$ is chosen in such a way that the
curvature perturbations vanish (flat slicing).  Here $N(t_f,t_i;{\vec x})$ is the local $e$-folding number, given by the integral of the local expansion along the worldline
${\vec x}={\rm const.}$ from $t=t_i$ to $t=t_f$.
The curvature perturbation evaluated on the uniform-density slicing, 
where the total energy density is spatially uniform, is denoted by $\zeta$.
Similarly we define the quantity $\zeta_i$ as the curvature perturbation evaluated on the slice
where the energy density of the $i$-th component becomes uniform ($\delta \rho_i(\vec x)=0 $).
Hereafter we take the uniform density slicing at $t=t_f$ in the last radiation dominated epoch
before relevant cosmological scales enter the horizon.
In the radiation dominated epoch the curvature perturbation is approximately given by
$\zeta \simeq \zeta_r$, where $\zeta_r$ denotes the curvature perturbation on a
slice where the energy density of the total radiation is spatially uniform.

Let us assume that CDM has isocurvature fluctuation $S$.
We allow CDM to be composed of multiple particle species each of which can have different origin.
Therefore, for instance,
some components of CDM may have isocurvature fluctuations while the remaining ones do not.

We define the CDM isocurvature peturbation as \cite{Wands:2000dp, Kawasaki:2008sn} \footnote{
The sign of $S$ given by Eq.~(\ref{Sdef}) is opposite to the one in \cite{Wands:2000dp}.}
\begin{eqnarray}
	S \equiv 3( \zeta_{\rm CDM}-\zeta_r ). \label{Sdef}
\end{eqnarray}
Here $\zeta_{\rm CDM}$ is the curvature perturbation 
on a slice where the CDM density becomes spatially uniform.
Then the total curvature perturbation in the matter dominated era is given by
\begin{equation}
	\zeta^{\rm MD} =  \zeta + \frac{1}{3}S, \label{relation1}
\end{equation}
where $\zeta$ is the total curvature perturbation in the radiation dominated era.

We consider a class of models in which 
$\zeta$ and $S$ originate from the quantum fluctuations $\{\delta \phi_a \}$
of light scalar fields $\{ \phi_a \}$ during inflation.
Note that the inflaton is also included in $\{ \phi_a \}$.
We can expand $\zeta$ and $S$ in terms of $\delta \phi_a$ as\footnote{
	The evolution of $\{ \phi_a \}$ is assumed to be smooth enough so that
	such expansions are justified.
}
\begin{gather}
	\zeta = N_a \delta \phi_a +\frac{1}{2}N_{ab}\delta \phi_a \delta \phi_b + \dots, 
	\label{zetaexp} \\
	S= S_{a} \delta \phi_a + \frac{1}{2}S_{ab}\delta \phi_a \delta \phi_b + \dots
	\label{Sexp},
\end{gather}
where $N_a \equiv \partial N/ \partial \phi_a$ and 
$S_a \equiv \partial S/ \partial \phi_a$ ,\footnote{
	Precisely speaking, $\{ \phi_a \}$ denote the field values during inflation,
	and they may take different values at the onset of oscillations.
	If the potentials of $\{ \phi_a \}$ are significantly deviated 
	from quadratic one, such differences 
	can be important for evaluating the non-Gaussianity \cite{Lyth:2005fi,Enqvist:2005pg}.
	In this paper we neglect such effects.
}
and summation over the repeated indices $a,b,\dots$ is implicitly taken.
Here we truncate the expansion at the second order and neglect higer order terms. 
For simplicity, we assume that the masses of $\{\phi_a\}$ are
negligibly small during and after inflation, and the fluctuations are independent to each other. Then
the correlation functions are given by the following form,
\begin{equation}
	\langle \delta \phi^a_{\vec k_1} \delta \phi^b_{\vec k_2} 
	\rangle\;=\;(2\pi)^3\,
	\delta(\vec k_1+\vec k_2)P_{\delta \phi}(k_1) \delta^{ab}
	\label{phiphi}
\end{equation}
with
\beq
	P_{\delta \phi}(k) \;\simeq\; \frac{H_{\rm inf}^2}{2k^3},
\eeq
where $k$ denotes the comoving wavenumber, and 
$H_{\rm inf}$ is the Hubble parameter during inflation.  For
later use, we also define the following:
\beq
	\Delta_{\delta \phi}^2 \;\equiv\; \frac{k^3}{2\pi^2} P_{\delta \phi}(k)
	\simeq \lrfp{H_{\rm inf}}{2\pi}{2}.
\eeq

Before formulating the bispectra from non-Gaussianities in the isocurvature perturbations, 
here we define the auto/cross-correlation functions (and their spectra) of 
the primordial curvature and isocurvature perturbations, $\zeta_{\vec k}$ and $S_{\vec k}$.
We define 
\begin{gather}
	\langle \zeta_{\vec k_1} \zeta_{\vec k_2} \rangle \equiv
	(2\pi)^3 \delta(\vec k_1+\vec k_2) P_\zeta(k_1), \\
	\langle \zeta_{\vec k_1} S_{\vec k_2} \rangle \equiv
	(2\pi)^3 \delta(\vec k_1+\vec k_2) P_{\zeta S}(k_1), \\
	\langle S_{\vec k_1} S_{\vec k_2} \rangle \equiv
	(2\pi)^3 \delta(\vec k_1+\vec k_2) P_{S}(k_1).
\end{gather}
Substituting (\ref{zetaexp}) and (\ref{Sexp}) into these equations, we obtain
\begin{gather}
	P_\zeta(k)=N_aN_a P_{\delta \phi}(k) + \frac{1}{2}N_{ab}N_{ab}
	\int \frac{d^3 \vec k'}{(2\pi)^3}P_{\delta \phi}(k')P_{\delta \phi}(|\vec k-\vec k'|),\\
	P_{\zeta S}(k)=N_aS_a P_{\delta \phi}(k) + \frac{1}{2}N_{ab}S_{ab}
	\int \frac{d^3 \vec k'}{(2\pi)^3}P_{\delta \phi}(k')P_{\delta \phi}(|\vec k-\vec k'|),\\
	P_S(k)=S_aS_a P_{\delta \phi}(k) + \frac{1}{2}S_{ab}S_{ab}
	\int \frac{d^3 \vec k'}{(2\pi)^3}P_{\delta \phi}(k')P_{\delta \phi}(|\vec k-\vec k'|).	
\end{gather}
After performing the integration, the spectra $P_\zeta$, $P_{\zeta S}$ and $P_S$ can be expressed as
\begin{gather}
	P_\zeta(k) \simeq [ N_a N_a + N_{ab}N_{ab}\Delta_{\delta \phi}^2 \ln (kL) ]
	P_{\delta \phi}(k), \\
	P_{\zeta S}(k) \simeq [ N_a S_a + N_{ab}S_{ab}\Delta_{\delta \phi}^2 \ln (kL) ] 
	P_{\delta \phi}(k), \\
	P_{S}(k) \simeq
	[ S_a S_a + S_{ab}S_{ab}\Delta_{\delta \phi}^2 \ln (kL) ] P_{\delta \phi}(k).
\end{gather}
Here we have introduced an infrared cutoff $L$, which is taken to be of order of the present Hubble
horizon scale \cite{Lyth:1991ub,Lyth:2007jh}.

We define a cross-correlation coefficient by $\gamma$,
\begin{equation}
	\gamma \equiv \frac{-P_{\zeta S}(k_0)}{\sqrt{P_\zeta(k_0)P_{S}(k_0) }}.
\end{equation}
Uncorrelated isocurvature perturbation corresponds to $\gamma = 0$ and 
totally (anti-)correlated one is $\gamma = (-)1$.
The initial condition for the structure formation is almost adiabatic, and 
the amplitude of isocurvature perturbations is now constrained from various cosmological observations.
For the uncorrelated case, the WMAP5 constraint is \cite{Komatsu:2008hk}
\begin{equation}
	\frac{P_{S}(k_0)}{P_\zeta(k_0)} \lesssim 0.190,  \label{Sbound}
\end{equation}
while for the totally anti-correlated case, the constraint is
\begin{equation}
	\frac{P_{S}(k_0)}{P_\zeta(k_0)}=
	\left ( \frac{P_{\zeta S}(k_0)}{P_\zeta(k_0)} \right )^2
	 \lesssim 0.0111.
\end{equation}
As we will see, these constraints give upper bounds on the non-linearity parameters
associated with the isocurvature perturbations.

\subsection{Non-Gaussianity from isocurvature perturbations}

Isocurvature perturbations 
must have negligible contribution to the power spectrum
of the total curvature perturbation from observations.
Therefore we can approximately obtain
\begin{equation}
	\langle \zeta_{\vec k_1}^{\rm MD} \zeta_{\vec k_2}^{\rm MD} \rangle \approx \langle \zeta_{\vec k_1} 	\zeta_{\vec k_2} \rangle 
	=(2\pi)^3\delta(\vec k_1+\vec k_2) P_\zeta(k_1),
\end{equation}
where
\begin{equation}
	P_\zeta(k) = \alpha_\zeta P_{\delta \phi}(k),
\end{equation}
and
\begin{equation}
	\alpha_\zeta = N_a N_a + N_{ab}N_{ab}\Delta_{\delta \phi}^2 \ln (kL). \label{po1}
\end{equation}
Meanwhile, the
isocurvature perturbation may significantly contribute to the three-point function of $\zeta^{\rm MD}$.
We define the bispectrum of $\zeta^{\rm MD}$,
$B^{\rm MD}_\zeta$ by the following equation,
\begin{eqnarray}
\langle \zeta_{ {\vec k_1}}^{\rm MD} \zeta_{ {\vec k_2}}^{\rm MD} \zeta_{ {\vec k_3}}^{\rm MD} \rangle
\equiv (2\pi)^3\delta(\vec k_1+\vec k_2+\vec k_3)~B^{\rm MD}_\zeta (k_1,k_2,k_3).
\end{eqnarray}
This contains four kind of terms, like 
$\langle \zeta\zeta\zeta\rangle, \langle \zeta\zeta S\rangle, 
\langle \zeta SS\rangle$ and $\langle SSS\rangle$. 
For example, focusing on the first contribution from $\langle \zeta\zeta\zeta\rangle$, 
the bispectrum includes the following terms,
\begin{equation}
\begin{split}
	B^{\rm MD}_\zeta (k_1,k_2,k_3)  \supset &
	N_aN_bN_{ab}\left[ P_{\delta \phi}(k_1) P_{\delta \phi}(k_2)+2~{\rm perms.} \right]\\
	&+N_{ab}N_{bc}N_{ca}\int \frac{d^3\vec k'}{(2\pi)^3}P_{\delta \phi}(k')
	P_{\delta \phi}(|\vec k_1-\vec k'|)P_{\delta \phi}(|\vec k_2-\vec k'|),
\end{split}
\end{equation}
and similar expressions hold for other three contributions.
Performing the integration,
we can express $B^{\rm MD}_\zeta$ in terms of the four combinations of the bispectrum of $\zeta$ and $S$,
\begin{eqnarray}
B^{\rm MD}_\zeta (k_1,k_2,k_3) \simeq
\left( \beta_{\zeta \zeta \zeta}+\frac{1}{3}\beta_{\zeta \zeta S}+\frac{1}{9}\beta_{\zeta SS}+\frac{1}{27}\beta_{SSS}\right) \left[ P_{\delta \phi}(k_1) P_{\delta \phi}(k_2)+2~{\rm perms.} \right],
\label{bi2}
\end{eqnarray}
where each term on the right hand side (RHS) is given by
\begin{eqnarray}
&&\beta_{\zeta \zeta \zeta}= N_a N_bN_{ab} + N_{ab}N_{bc}N_{ca}\Delta_{\delta \phi}^2 \ln (k_bL), \label{beta1}  \\
&&\beta_{\zeta \zeta S}= N_a N_bS_{ab} + 2N_{ab}N_a S_b +3N_{ab}N_{bc}S_{ca}\Delta_{\delta \phi}^2 \ln (k_bL), \label{beta2} \\
&&\beta_{\zeta S S}=S_a S_bN_{ab} + 2S_{ab}S_a N_b +3S_{ab}S_{bc}N_{ca}\Delta_{\delta \phi}^2 \ln (k_bL), \label{beta3}  \\
&&\beta_{SSS}=S_a S_bS_{ab} + S_{ab}S_{bc}S_{ca}\Delta_{\delta \phi}^2 \ln (k_bL),
\label{beta4}  
\end{eqnarray}
in a squeezed configuration 
that one of the three wave vectors is much smaller than the other two
(e.g. $k_1\ll k_2,k_3$) and we have defined $k_b \equiv \min \{ k_1, k_2, k_3 \}$.

In many literatures concerning the non-Gaussianity of the primordial fluctuations, 
where the adiabaticity is implicitly assumed,
the magnitude of the non-Gaussianity of the curvature perturbation is conventionally
parametrized by the so-called non-linearity parameter $f_{\rm NL}$ which is defined
by the ratio of the bispectrum to the square of the power spectrum.
Since a single parameter is not enough to parametrize the non-Gaussianity in the presence 
of both the adiabatic and isocurvature perturbations because of their different effects
on matter spectrum or temperature anisotropy of the CMB,
we define four types of non-linerity parameters as following,
\begin{eqnarray}
	&&\frac{6}{5} f_{\rm NL}^{(\rm adi)} = \alpha_\zeta^{-2} \beta_{\zeta \zeta \zeta},~~~~~
	\frac{6}{5} f_{\rm NL}^{(\rm cor1)} = \frac{1}{3}\alpha_\zeta^{-2} \beta_{\zeta \zeta S},~~
	\nonumber \\
	&&\frac{6}{5} f_{\rm NL}^{(\rm cor2)} =\frac{1}{9} \alpha_\zeta^{-2} \beta_{\zeta SS},~~
	\frac{6}{5} f_{\rm NL}^{(\rm iso)} = \frac{1}{27} \alpha_\zeta^{-2} \beta_{SSS},
\end{eqnarray}
These formulae reproduce the known results for the curvaton, ungaussiton and axion, 
if only one of these non-linearity parameters exists, as we will see.
However, in general, all the four non-linearity parameters can be concomitant
and their effects on the temperature anisotropy 
have not been investigated in the previous literatures.

\section{Applications} \label{sec:app}

In this section,
as an application of the formalism given in the previous section,
we consider some simple models where the inflaton $\phi$ and another light field
$\sigma$ contributes to the adiabatic and isocurvature perturbations.
We neglect the non-Gaussianity generated by the inflaton itself (i.e. we set $N_{\phi \phi}=0$).

\subsection{Simple examples}

First we demonstrate that our formulae given in the previous section correctly 
reproduce known results for some simple models.

\subsubsection{Curvaton model with no CDM isocurvature perturvation}

If all the CDM is generated after the decay of the curvaton,
no isocurvature perturbations are generated.
Assuming $N_\sigma \gg N_\phi$, i.e.,
the curvature perturbations are dominantly generated by the curvaton,
we have
\begin{equation}
	\frac{6}{5}f_{\rm NL}^{(\rm adi)}=\frac{1}{N_\sigma^4}
	[ N_\sigma^2 N_{\sigma \sigma}
	+ N_{\sigma \sigma}^3 \Delta_{\delta \sigma}^2 \ln(k_bL) ].
\end{equation}
This is the standard result for the curvaton model \cite{Lyth:2005fi}.
But in general cases, curvaton models predict the existence of correlated CDM isocurvature 
perturbation, and it can significantly modify the feature of non-Gaussianity in the CMB anisotropy,
as we will see later.

\subsubsection{Ungaussiton model with no CDM isocurvature perturvation}

Similar to the curvaton case, but $\sigma$ is assumed to only affect the bispectrum of the
curvature perturbation, and have only small effect on the power spectrum.
In the limit $N_\sigma \ll N_\phi$, there are no isocurvature perturbations 
if all the CDM is generated from the inflaton decay products.
In this case we obtain
\begin{equation}
	\frac{6}{5}f_{\rm NL}^{(\rm adi)}=\frac{1}{N_\phi^4}[ N_\sigma^2 N_{\sigma \sigma}
	+ N_{\sigma \sigma}^3 \Delta_{\delta \sigma}^2 \ln(k_bL) ],
\end{equation}
as shown in \cite{Suyama:2008nt,Lyth:2006gd}.

\subsubsection{Axion}

Let us assume that $\sigma$ is stable and contributes to some fraction of the CDM.
The axion ($a$) \cite{Peccei:1977hh}
is one of the well motivated candidates of such a scalar field.
Since the axion has only negligible energy density in the radiation dominated phase, we have
$N_\sigma \sim 0 \ll S_\sigma$ and $\beta_{\zeta  \zeta S}\sim \beta_{\zeta SS}\sim 0$,
corresponding to an uncorrelated isocurvature perturbation.
Thus we obtain
\begin{equation}
	\frac{6}{5}f_{\rm NL}^{(\rm iso)}=\frac{1}{27N_\phi^4}[ S_\sigma^2 S_{\sigma \sigma}
	+ S_{\sigma \sigma}^3 \Delta_{\delta \sigma}^2 \ln(k_bL) ].
\end{equation}
This coincides with our previous result \cite{Kawasaki:2008sn}.

\subsection{General curvaton model}\label{gcm}

Let us consider a case that both radiation and CDM 
originate from the decay of both the inflaton ($\phi$) and the curvaton ($\sigma$).\footnote{
	The following discussions do not depend on which of them dominates the total
	curvature perturbation, but conveniently we call $\sigma$ `curvaton' in both cases.
} 
The radiation and the CDM produced from the inflaton have
the same fluctuations,
i.e. the energy density of the radiation becomes spatially uniform
on the slice where that of the CDM becomes spatially uniform.
We denote the curvature perturbation on this slice by $\zeta_\phi$.
Similarly, we define $\zeta_\sigma$ as the curvature perturbation on a slicing on which 
the radiation and the CDM produced by the curvaton become spatially uniform.

\subsubsection{CDM from the inflaton and direct decay of the curvaton}

Let us first consider the case that some fraction of the CDM 
are produced directly from the decay of the curvaton.
We assume that the universe is dominated by the radiation at the curvaton decays,
and so we can approximate the total curvature perturbation as $\zeta \simeq \zeta_r$.
Also we assume that the curvaton decays instantaneously when $H=\Gamma_\sigma$,
where $\Gamma_\sigma$ is decay rate of the curvaton.
Then taking a uniform density slicing just before the curvaton decays,
we have a relation given by
\begin{eqnarray}
\rho_r^\phi ( {\vec x})+f \rho_\sigma ({\vec x})={\rho_r}(\vec x), \label{ddc1}
\end{eqnarray}
where $f$ is the fraction of the curvaton energy density that transfers
to the radiation.
Because $\rho_r^\phi$ and $\rho_\sigma$ have different origin,
they are inhomogeneous in general on this slice.
Using equations $\rho_r^\phi=e^{4(\zeta_\phi-\zeta)} {\bar \rho_r^\phi},~\rho_\sigma=e^{3(\zeta_\sigma-\zeta)} {\bar \rho_\sigma}$,
$\rho_r=\bar \rho_r e^{4(\zeta_r-\zeta)}$ and $\zeta \simeq \zeta_r$,
Eq.~(\ref{ddc1}) can be written as\footnote{
	Precisely speaking,  when the $\sigma$
	decays into radiation, there is a subtlety in the definition of a slicing 
	on which the energy density of $\sigma$ is spatially uniform.
	The equations in the text are valid if we interpret 
 	the relation between $\rho_\sigma$ on the uniform density slicing and
	its background value as the definition of $\zeta_\sigma$.
	This is because the energy density of $\sigma$
	on the uniform density slicing is well defined even at moment $\sigma$ decays.
}
\begin{eqnarray}
(1-\epsilon_r) e^{4(\zeta_\phi-\zeta_r)}+\epsilon_r e^{3(\zeta_\sigma-\zeta_r)}=1, \label{ddc2}
\end{eqnarray}
where $\epsilon_r \equiv f {\bar \rho_\sigma} / {\bar \rho_r} |_{\rm decay}$
denotes the fraction of the radiation produced from the curvaton.
Hence $1-\epsilon_r$ is the fraction of the radiation from the inflaton.

In the same way,
we obtain the similar equation to Eq.~(\ref{ddc2}) for the CDM as
\begin{eqnarray}
(1-\epsilon_{\rm CDM}) e^{3(\zeta_\phi-\zeta_{\rm CDM})}+\epsilon_{\rm CDM} e^{3(\zeta_\sigma-\zeta_{\rm CDM})}=1, \label{ddc3}
\end{eqnarray}
where $\epsilon_{\rm CDM} \equiv (1-f) {\bar \rho_\sigma} / {\bar \rho_{\rm CDM}} |_{\rm decay}$
denotes the fraction of the CDM produced from the curvaton.
Eqs.~(\ref{ddc2}) and (\ref{ddc3}) give the fully non-linear
relations between $(\zeta_r, \zeta_{\rm CDM})$ and $(\zeta_\phi, \zeta_\sigma)$
under the sudden decay approximation.

Expanding the perturbation variables in Eqs.~(\ref{ddc2}) and (\ref{ddc3}) up to
the second order,
we can explicitly express $\zeta_r$ and $\zeta_{\rm CDM}$ in terms of $\zeta_\phi$ and 
$\zeta_\sigma$ as
\begin{eqnarray}
&&\zeta_r=(1-R)\zeta_\phi+R\zeta_\sigma+\frac{1}{2}R(1-R)(3+R) {( \zeta_\phi-\zeta_\sigma )}^2, \\
&&\zeta_{\rm CDM}=(1-\epsilon_{\rm CDM}) \zeta_\phi+\epsilon_{\rm CDM} \zeta_\sigma+\frac{3}{2} \epsilon_{\rm CDM} (1-\epsilon_{\rm CDM}) {(\zeta_\phi-\zeta_\sigma)}^2,
\end{eqnarray}
where we have introduced a new parameter $R$ defined by
\begin{eqnarray}
R=\frac{3 \epsilon_r}{4-\epsilon_r}.
\end{eqnarray}

Meanwhile,
the curvaton energy density on the uniform density slice can be written as
$\rho_\sigma({\vec x})=e^{3 (\zeta_\sigma-\zeta)} {\bar \rho_\sigma}$.
If we take this slice just after the curvaton starts its oscillation,
the universe is dominated by the radiation produced from the inflaton\footnote{
	We assume that vacuum expectation value of the curvaton during inflation is much smaller than
	$M_{\rm pl}$, where $M_{\rm pl} =2.4\times 10^{18}$~GeV is the reduced Planck scale.
	In this case,
	the fraction of the curvaton energy density at the time when the curvaton begins 
	its oscillations is roughly given by $\rho_\sigma/(M_{\rm pl}^2 m_\sigma^2) \simeq \sigma^2/M_{\rm pl}^2 \ll 1$.
	Hence the curvaton is subdominant.
	Note also that it does not matter whether the inflaton has already decayed or not when the curvaton starts
	its oscillations.} 
and hence $\zeta=\zeta_\phi$.
Denoting the density contrast of the curvaton energy density on the uniform density slicing 
by $\delta_\sigma \equiv (\rho_\sigma(\vec x)-\bar \rho_\sigma)/\bar \rho_\sigma$, we find 
\begin{eqnarray}
\zeta_\sigma=\zeta_\phi+\frac{1}{3} \log (1+\delta_\sigma).
\end{eqnarray}
Up to the second order in the curvaton fluctuation,
the above equation can be written as
\begin{eqnarray}
\zeta_\sigma=\zeta_\phi+\frac{1}{3} \left( \delta_\sigma -\frac{1}{2} \delta_\sigma^2 \right).
\end{eqnarray}
Then,
using the standard formula,
\begin{eqnarray}
\zeta_\phi=\frac{1}{M_{\rm pl}^2} \frac{V}{V_\phi} \delta \phi+\frac{1}{2M_{\rm pl}^2} \left( 1-\frac{V V_{\phi \phi}}{V_\phi^2} \right) {(\delta \phi)}^2,
\end{eqnarray}
where $V$ is the potential of the inflaton, $V_{\phi}$ and $V_{\phi \phi}$ are its first and second
derivative with respective to $\phi$.
Noting that $\delta_\sigma=2\delta \sigma/\sigma+{(\delta \sigma)}^2/\sigma^2$,
we can express $\zeta_r$ and $\zeta_\sigma$ in terms of $\delta \phi$ and $\delta \sigma$ as
\begin{eqnarray}
&&\zeta_r=\frac{1}{M_{\rm pl}^2} \frac{V}{V_\phi} \delta \phi+\frac{1}{2M_{\rm pl}^2} \left( 1-\frac{V V_{\phi \phi}}{V_\phi^2} \right) {(\delta \phi)}^2+\frac{2R}{3\sigma} \delta \sigma+\frac{R}{9 \sigma^2} (3-4R-2R^2) {( \delta \sigma)}^2, \label{zr1} \\
&&\zeta_{\rm CDM}=\frac{1}{M_{\rm pl}^2} \frac{V}{V_\phi} \delta \phi+\frac{1}{2M_{\rm pl}^2} \left( 1-\frac{V V_{\phi \phi}}{V_\phi^2} \right) {(\delta \phi)}^2 \nonumber \\  
&&{\hskip 5cm}+\frac{2\epsilon_{\rm CDM}}{3\sigma} \delta \sigma+\frac{\epsilon_{\rm CDM} (1-2 \epsilon_{\rm CDM})}{3 \sigma^2}{( \delta \sigma)}^2. \label{zc1}
\end{eqnarray}
As expected,
no cross terms of $\delta \phi$ and $\delta \sigma$ appear in the final expressions.
From these results,
we can immediately read the expansion coefficients as 
\begin{eqnarray}
&&N_\phi=\frac{1}{M_{\rm pl}^2} \frac{V}{V_\phi} ,~~~N_\sigma=\frac{2R}{3\sigma},\\
&&N_{\phi \phi}=\frac{1}{M_{\rm pl}^2} \left( 1-\frac{V V_{\phi \phi}}{V_\phi^2} \right),~~~N_{\sigma \sigma}=\frac{2R}{9 \sigma^2} (3-4R-2R^2), \\
&&S_\phi=S_{\phi \phi}=0, \\
&&S_\sigma=2(\epsilon_{\rm CDM}-R)\frac{1}{\sigma},~~~S_{\sigma \sigma}=2 \bigg[ \epsilon_{\rm CDM} (1-2 \epsilon_{\rm CDM})-\frac{R}{3}(3-4R-2R^2) \bigg] \frac{1}{\sigma^2}.
\end{eqnarray}
We can see that except for the trivial limit $R=0$ and $\epsilon_{\rm CDM} = 0$,
isocurvature perturbation vanishes only when $R=1$ and $\epsilon_{\rm CDM}=1$,
corresponding to the case that the curvaton dominates the universe before it decays and 
all CDM is generated by the curvaton decay itself.
Using this result,
we can derive the following relations,
\begin{eqnarray}
&&\frac{6}{5}f_{\rm NL}^{(\rm adi)}=\frac{1}{2(1+p)^2R}(3-4R-2R^2), \\
&&\frac{6}{5}f_{\rm NL}^{(\rm cor1)}
=\frac{3}{2(1+p)^2R^2}\bigg[ \epsilon_{\rm CDM}(1-2\epsilon_{\rm CDM})+\left( \frac{2}{3}\epsilon_{\rm CDM}-R \right) \left( 3-4R-2R^2\right) \bigg], \\
&&\frac{6}{5}f_{\rm NL}^{(\rm cor2)}
=\frac{3}{2(1+p)^2R^3} (\epsilon_{\rm CDM}-R) \nonumber \\
&&{\hskip 2cm}\times \bigg[ \epsilon_{\rm CDM} \left(3-\frac{4R}{3}-\frac{2R^2}{3}-4 \epsilon_{\rm CDM} \right)-R(3-4R-2R^2) \bigg], \\
&&\frac{6}{5}f_{\rm NL}^{(\rm iso)}
=\frac{3}{2(1+p)^2R^4} {(\epsilon_{\rm CDM}-R)}^2 \bigg[ \epsilon_{\rm CDM}(1-2\epsilon_{\rm CDM})-\frac{R}{3} (3-4R-2R^2) \bigg],
\end{eqnarray}
when the fluctuation of the curvaton is dominated by the linear part, that is, 
$\delta_\sigma \sim 2\delta \sigma/\sigma$.
Here $p \equiv N_\phi^2/N_\sigma^2=9\sigma^2 V^2/(4M_{\rm pl}^4 V_\phi^2 R^2)$ represents the 
ratio of the inflaton contribution to the total curvaure perturbation and that of the curvaton.
The limit $p=0$ corresponds to the standard curvaton scenario.
In deriving these results,
we have neglected the non-Gaussianity from the inflaton fluctuation because it gives
$f_{\rm NL}$ a value of order of the slow-roll parameters which are much smaller than unity
\cite{Acquaviva:2002ud,Maldacena:2002vr,Seery:2005wm,Yokoyama:2007uu}.

Truncating the perturbative expansion in Eqs.~(\ref{zr1}) and (\ref{zc1})
at the linear order,
we find 
\begin{eqnarray}
&&\frac{P_S}{P_\zeta}=\frac{S_\sigma^2}{N_\phi^2+N_\sigma^2}
=\frac{9 {(\epsilon_{\rm CDM}-R)}^2}{ (1+p) R^2}, \\
&&\gamma=-\frac{P_{\zeta S}}{\sqrt{ P_\zeta P_S}}=-\frac{1}{\sqrt{1+p}}.
\end{eqnarray}
Note that
the cross-correlation parameter $\gamma$ is reduced to $-1$ in the standard curvaton scenario.

Let us suppose $p=0$ for simplicity and consider the case of the large non-Gaussianity $f_{\rm NL} \gtrsim 10$.
To have the large $f_{\rm NL}$,
$R$ must be much smaller than unity.
From the isocurvature constraints,
$\epsilon_{\rm CDM}$ must be very close to $R$ and hence $\epsilon_{\rm CDM}$
is also a small quantity.
Then the term $\beta_{\zeta \zeta \zeta}$ is the largest
and the second largest term $\beta_{\zeta \zeta S}$ is suppressed by $\epsilon_{\rm CDM}$
compared to $\beta_{\zeta \zeta \zeta}$.
Other terms are more suppressed by the power of $\epsilon_{\rm CDM}$.
Hence the leading contribution to $f_{\rm NL}$ from the isocurvature perturbation
comes from the term $\beta_{\zeta \zeta S}$.

\subsubsection{CDM from the inflaton and thermal bath after the curvaton decay}  \label{sec:TPCDM}

Let us next consider a case that in addition to the CDM (denoted by X) 
produced from the inflaton decay,
CDM (denoted by Y) are also produced thermally during the radiation dominated era
after the curvaton decay.
We assume that all the curvaton energy density decays into the radiation.

Similar to the previous case, taking the uniform density slice just before curvaton decays, we obtain
\begin{eqnarray}
(1-\epsilon_r) e^{4(\zeta_\phi-\zeta_r)}+\epsilon_r e^{3(\zeta_\sigma-\zeta_r)}=1. \label{tcc2}
\end{eqnarray}
On the other hand,
Taking the uniform density slice after the total CDM is created,
we have the following relation,
\begin{eqnarray}
\epsilon_X e^{3(\zeta_X-\zeta_{\rm CDM})}+\epsilon_Y e^{3(\zeta_Y-\zeta_{\rm CDM})}=1, \label{tcc1}
\end{eqnarray}
where $\epsilon_X/\epsilon_Y$ is the fraction of $X/Y$ in the CDM.
Because CDM is assumed to be composed of $X$ and $Y$,
$\epsilon_X+\epsilon_Y=1$ must be satisfied.
X is produced from the decay product of the inflaton.
Hence $\zeta_X=\zeta_\phi$.
Meanwhile,
Y is produced from the radiation which originate from both
the inflaton and the curvaton.
Hence $\zeta_Y$ is equal to the curvature perturbation on the slice
where the total radiation energy density becomes homogeneous, $\zeta_Y = \zeta_r$.
Eqs.~(\ref{tcc1}) and (\ref{tcc2}) give the fully non-linear relation
in the case of the thermally produced CDM.
Expanding these equations up to the second order, we obtain
\begin{eqnarray}
&&\zeta_r=(1-R)\zeta_\phi+R\zeta_\sigma+\frac{1}{2}R(1-R)(3+R) {( \zeta_\phi-\zeta_\sigma )}^2, \\
&&\zeta_{\rm CDM}=\zeta_\phi-R(1-\epsilon_{X})(\zeta_\phi-\zeta_\sigma) \nonumber \\
&&{\hskip 2cm}+\frac{1}{2} R (1-\epsilon_{X}) \left[ 3R\epsilon_X+(1-R)(3+R)\right ] {(\zeta_\phi-\zeta_\sigma)}^2.
\end{eqnarray}

By doing the same procedures as in the previous subsection,
we arrive at
\begin{equation}
	S_\sigma = -\frac{2R\epsilon_X}{\sigma}, ~~~
	S_{\sigma \sigma} = -\frac{2R\epsilon_X}{3}\left[ 3-4R-2R^2 - 6R(1-\epsilon_X)
	\right] \frac{1}{\sigma^2}.
\end{equation}
From this we can immediately see that the isocurvature perturbation vanishes if $\epsilon_X=0$, i.e.,
all CDM is generated thermally after the curvaton decays, as expected.
Other quantities ($N_\phi, N_\sigma,\dots$) are same as those obtained in the previous subsection.
Thus we can calculate non-linearity parameters as
\begin{eqnarray}
&& \frac{6}{5}f_{\rm NL}^{(\rm adi)}
=\frac{1}{2(1+p)^2R}(3-4R-2R^2), \\ \label{eq:fNL1}
&& \frac{6}{5}f_{\rm NL}^{(\rm cor1)}
=-\frac{3 \epsilon_X}{2(1+p)^2R} \bigg[ 3-4R-2R^2-2R(1-\epsilon_X) \bigg], \\ \label{eq:fNL2}
&& \frac{6}{5}f_{\rm NL}^{(\rm cor2)}
=\frac{3\epsilon_X^2}{2(1+p)^2R}\bigg[ 3-4R-2R^2-4R(1-\epsilon_X) \bigg], \\ \label{eq:fNL3}
&& \frac{6}{5}f_{\rm NL}^{(\rm iso)}
=-\frac{\epsilon_X^3}{2(1+p)^2R} \bigg[ 3-4R-2R^2-6R(1-\epsilon_X) \bigg], \\ \label{eq:fNL4}
&&\frac{P_S}{P_\zeta}=\frac{9\epsilon_X^2}{1+p}, \\
&&\gamma=-\frac{1}{\sqrt{1+p}}.
\end{eqnarray}

As the simplest case,
let us assume $p=0$.
Then, 
from the isocurvature constraints,
we have a upper bound on $\epsilon_X$ as
\begin{eqnarray}
	\epsilon_X \lesssim 0.035. \label{constraint_cor}
\end{eqnarray}
We find $\epsilon_X$ must be much smaller than unity.
In this case,
the leading contribution to the non-Gaussianity from the isocurvature perturbation
comes from $f_{\rm NL}^{(\rm cor1)}$ because $f_{\rm NL}^{(\rm cor2)}$ and $f_{\rm NL}^{(\rm iso)}$
are more suppressed by the power of $\epsilon_X$.

Through the analyses in the previous and this subsection,
we have found that in both cases the bispectrum of the adiabatic perturbation 
dominantly contribute to the non-Gaussianity while the leading contribution
from the isocurvature perturbation comes from $f_{\rm NL}^{(\rm cor1)}$ which
are suppressed by the small parameter.
But this does not necessarily mean the isocurvature contribution is irrelevant.
Since $f_{\rm NL}^{(\rm cor1)}$ represents the contribution from the 
non-Gaussian fluctuation of the correlated isocurvature perturvation between CDM and photon,
it has more drastic effects on large-scale CMB anisotropy.
Actually it was pointed out in \cite{Kawasaki:2008sn} that
an isocurvature type non-Gaussianity of the large scale temperature anisotropy 
can be about 100 times larger than
the usual adiabatic case, even if $f_{\rm NL}^{(\rm adi)}=f_{\rm NL}^{(\rm iso)}$.
Thus it may be the case that the non-Gaussianity in the curvaton model is dominated by the
isocurvature type, while satisfying the constraint on the magnitude of the isocurvature fluctuation
from the power spectrum.
We need numerical calculation to quantify the consequences of such mixed
adiabatic and isocurvature non-Gaussianity.

\subsubsection{A `realistic' example} \label{sec:example}

Here some particle physics motivated examples of the curvaton model are exhibited,
and we show explicitly that it is natural to expect the existence of a correlated isocurvature perturbation
in realistic curvaton models, if a large non-Gaussianity is generated by the curvaton.

First note that $R$ is given by
\begin{equation}
	R \simeq \left \{ \begin{array}{ll}
	\displaystyle \frac{\sigma_i^2}{4M_{\rm pl}^2}\frac{T_R}{T_\sigma}
	~~&{\rm for}~~~m_\sigma > \Gamma_\phi \\
	\displaystyle \frac{\sigma_i^2}{4M_{\rm pl}^2}\frac{T_{\rm osc}}{T_\sigma}
	~~&{\rm for}~~~m_\sigma < \Gamma_\phi
	\end{array} \right. ,
\end{equation}
where $m_\sigma$ is the mass of the curvaton, 
$\Gamma_\phi$ is the decay rate of the inflaton,
$T_\sigma$ is the decay temperature of $\sigma$,
$T_R$ is the reheating temperature after inflation defined by 
$T_R=(10/\pi^2 g_*)^{1/4}\sqrt{\Gamma_\phi M_{\rm pl}}$ with the effective relativistic degrees of
freedom $g_*$,
$T_{\rm osc}$ is the temperature at which the curvaton begins to oscillate defined by
$T_{\rm osc}=(10/\pi^2 g_*)^{1/4}\sqrt{m_\sigma M_{\rm pl}}$,
and $\sigma_i$ is the initial amplitude of the curvaton, which is assumed to be
much smaller than $M_{\rm pl}$.\footnote{
	Note that in order that the curvaton can generate the observed magnitude of the
	 density perturbation, $R$ cannot be smaller than $10^{-5}$.
}
In order to generate sizable non-Gaussianity, $R$ should be around 0.1.

In a SUSY theory, there exist many scalar fields and some of them may remain light during inflation.
These light scalars are candidates for the curvaton \cite{Hamaguchi:2003dc}.
Good examples are the right-handed sneutrino \cite{Moroi:2002vx}, 
flat direction in the minimal supersymmetric standard model \cite{Enqvist:2003mr,Riotto:2008gs},
scalar partner of the axion (saxion) \cite{Rajagopal:1990yx,Dimopoulos:2003ii}, and so on.
We assume $p=0$ and that the curvaton decays 
before the LSP freezes out $T_\sigma \gtrsim m_{\rm LSP}/20$,
where $m_{\rm LSP}$ denotes the mass of the lightest supersymmetric particle (LSP),
which is assumed to be the lightest neutralino.
In that case, the LSPs are produced in thermal bath after the curvaton decays,
and they can be the dark matter if their annihilation cross section takes an appropriate value
($\langle \sigma v \rangle \sim 3\times 10^{-26}~{\rm cm^3 s^{-1}}$).
These LSPs do not have isocurvature perturbation.
However, in general, LSPs are also produced nonthermally by the gravitino decay,
and gravitinos are produced at the reheating epoch after inflation
\cite{Bolz:2000fu,Kawasaki:2006gs}.
Thus energy density of the gravitino fluctuates in the same way as that of the inflaton,
so does the LSP directly created by the gravitino decay.\footnote{
	Since gravitinos decay well after the freezeout of the LSP, 
	nonthermally produced LSPs by the gravitino decay remain decoupled with thermal bath.
}
Thus those nonthermal LSPs have (correlated) isocurvature perturbations.
This is exactly the case of Sec.~\ref{sec:TPCDM}.
Here $\epsilon_X$ reads the fraction of the nonthermally produced LSP to the total LSP abundance.
The gravitino number-to-entropy ratio $(Y_{3/2})$ is given by \cite{Bolz:2000fu,Kawasaki:2004yh}
\begin{equation}
	Y_{3/2} \sim 2\times 10^{-12}\left ( \frac{T_R}{10^{10}~{\rm GeV}} \right )
	\left ( 1+\frac{m_{\tilde g}^2}{3m_{3/2}^2} \right ),
\end{equation}
where $m_{\tilde g}$ and $m_{3/2}$ denote the mass of the gluino and gravitino, respectively.
Thus the fraction of the nonthermally produced LSP in the total dark matter abundance is
estimated as
\begin{equation}
	\epsilon_X \sim 5\times 10^{-3}\left ( \frac{m_{\rm LSP}}{1~{\rm TeV}} \right )
	\left ( \frac{T_R}{10^{7}~{\rm GeV}} \right ).
\end{equation}
Thus in order to satisfy the constraint (\ref{constraint_cor}), we have upper bound on $T_R$ as
\begin{equation}
	T_R \lesssim 7\times 10^7~{\rm GeV}\left ( \frac{1~{\rm TeV}}{m_{\rm LSP}} \right ),
\end{equation}
which is marginally consistent with an upper bound from the cosmological gravitino problem
\cite{Kawasaki:2004yh}.
On the other hand, $T_R$ cannot be much smaller than this value,
since otherwise $R$ becomes too small, yielding too large non-Gaussianity,
which conflicts with observations.

Therefore, it seems natural to expect that some fraction of the LSP dark matter has
different origin, which inevitably involves isocurvature fluctuations.
In such a situation, an analysis usually done in many literatures assuming the existence of only an
adiabatic non-Gaussianity is not valid.
Instead we need to carefully study the effect of (correlated) isocurvature perturbations
on the resulting non-Gaussian features in the CMB anisotropy.

\section{CMB temperature anisotropy} \label{sec:CMB}

In this section we investigate the effects of the non-linear isocurvature perturbation 
on the CMB anisotropy, following the notations used in \cite{Bartolo:2004if,Komatsu:2001rj}.
Since the adiabatic and isocurvature perturbations have quite different properties 
regarding their imprints on the CMB anisotropy,
we need to correctly evaluate the bispectrum of the temperature anisotropy in the presence of
both adiabatic and isocurvature non-Gaussianity.

From temperature anisotropies originated from the adiabatic and isocurvature perturbations
$\Delta T^{(\rm adi)}(\vec n)$ and $\Delta T^{(\rm iso)}(\vec n)$
for a given direction $\vec n$, we define $a_{\ell m}$ by
\begin{eqnarray}
	a_{\ell m}\;=
	\;\int d{\vec n} ~\left [ \frac{\Delta T^{\rm (adi)}({\vec n})}{T} +
	\frac{\Delta T^{\rm (iso)}({\vec n})}{T}\right ]\, Y_{\ell m}^* ({\vec n}). 
\label{tem1}
\end{eqnarray}
Transfer functions are defined by
\begin{equation}
	\Theta^{\rm (adi)}_\ell({\vec k}) \;\equiv\; g^{\rm (adi)}_{T \ell}(k)\zeta_{\vec k},
\end{equation}
and
\begin{equation}
	\Theta^{\rm (iso)}_\ell({\vec k}) \;\equiv\; g^{\rm (iso)}_{T \ell}(k)S_{\vec k},\label{eq:deftrans}
\end{equation}
where $\Theta^{\rm (adi/iso)}_\ell({\vec k})$ is the multipole moment of CMB temperature anisotropy
from the adiabatic/isocurvature perturbation:
\begin{equation}
	\frac{\Delta T^{\rm (adi/iso)}({\vec n})}{T}\;=\;\int \frac{d^3k}{(2\pi)^3}
	\sum_\ell i^\ell (2\ell+1)\Theta^{\rm (adi/iso)}_\ell({\vec k})
	P_\ell({\hat {\vec k}}\cdot{\vec n}). \label{eq:phbf}
\end{equation}
Here $P_\ell$'s are the Legendre polynomials.
From these equations,
multipole moments can be divided as $a_{\ell m}=a_{\ell m}^{\rm (adi)}+a_{\ell m}^{\rm (iso)}$ where
\begin{eqnarray}
&&a_{\ell m}^{\rm (adi)}\;=\;4\pi i^\ell \int \frac{d^3 k}{ {(2\pi)}^3 }\, 
g_{T \ell}^{\rm (adi)} (k) \,Y^*_{\ell m}({\hat {\vec k}}) \,\zeta_{\vec k}, \label{tem1}\\
&&a_{\ell m}^{\rm (iso)}\;=\;4\pi i^\ell \int \frac{d^3 k}{ {(2\pi)}^3 }\, 
g_{T \ell}^{\rm (iso)} (k) \,Y^*_{\ell m}({\hat {\vec k}}) \,S_{\vec k}. \label{tem2}
\end{eqnarray}

The anglar power spectrum of $a_{\ell m}$ is calculated as
\beq
	\langle a_{\ell m} a_{\ell' m'}^{*} \rangle\; \equiv\;
	\left[ C_\ell^{\rm (adi)} + 2C_\ell^{\rm (cor)} + C_\ell^{\rm (iso)} \right ]
	\delta_{\ell \ell'} \delta_{mm'}.
\eeq
Using (\ref{tem2}), we obtain
\begin{eqnarray}
	&&C_\ell^{\rm (adi)}\;=\;\frac{2}{\pi} \int_0^\infty dk~k^2 \left( g_{T \ell}^{\rm (adi)}(k) \right)^2 
	P_{\zeta} (k), \\
	&&C_\ell^{\rm (cor)}\;=\;\frac{2}{\pi} \int_0^\infty dk~k^2 g_{T \ell}^{\rm (adi)}(k)  
	g_{T \ell}^{\rm (iso)}(k) P_{\zeta S} (k),  \label{Clcor} \\	
	&&C_\ell^{\rm (iso)}\;=\;\frac{2}{\pi} \int_0^\infty dk~k^2 {\left( g_{T \ell}^{\rm (iso)}(k) \right)}^2 
	P_S (k).			\label{tem3} 
\end{eqnarray} 

Similarly, the angular bispectrum of $a_{\ell m}$ is defined by
\begin{eqnarray}
	\langle a_{\ell_1 m_1}a_{\ell_2 m_2}a_{\ell_3 m_3}\rangle 
	\;\equiv\; B^{m_1 m_2 m_3}_{\ell_1 \ell_2 \ell_3}. \label{tem4}
\end{eqnarray}
Statistical isotropy divides the angular bispectrum into
the following form,
\begin{eqnarray}
	B^{ m_1 m_2 m_3}_{\ell_1 \ell_2 \ell_3}\;=
	\;{\cal G}^{m_1 m_2 m_3}_{\ell_1 \ell_2 \ell_3} b_{\ell_1 \ell_2 \ell_3}. 
	\label{tem5}
\end{eqnarray}
Here ${\cal G}^{m_1 m_2 m_3}_{\ell_1 \ell_2 \ell_3}\equiv \int d{\vec n}~Y_{\ell_1 m_1}({\vec n})Y_{\ell_2 m_2}({\vec n})Y_{\ell_3 m_3}({\vec n})$ (Gaunt integral) and $b_{\ell_1 \ell_2 \ell_3} $ is the reduced bispectrum, on which we will focus in the following.

Substituting (\ref{tem1}) and (\ref{tem2}) into (\ref{tem4}), we obtain
\begin{eqnarray}
	&&b_{\ell_1 \ell_2 \ell_3}= \frac{8}{\pi^3}\int_0^\infty dr~r^2 \int_0^\infty dk_1~k_1^2
	\int_0^\infty dk_2~k_2^2 \int_0^\infty dk_3~k_3^2 
	j_{\ell_1}(k_1r) j_{\ell_2}(k_2r)j_{\ell_3}(k_3r) \nonumber \\
	&&\hspace{15mm} \times \left[  g_{T \ell_1}^{\rm (adi)}(k_1)g_{T \ell_2}^{\rm (adi)} (k_2) 
	g_{T \ell_3}^{\rm (adi)} (k_3) \beta_{\zeta \zeta \zeta} (k_1,k_2,k_3) \right. \nonumber\\
	&&\hspace{20mm}
	+ \{g_{T \ell_1}^{\rm (adi)}(k_1)g_{T \ell_2}^{\rm (adi)} (k_2) 
	g_{T \ell_3}^{\rm (iso)} (k_3)+\mbox{(cyclic with $\ell$'s)} \}
	\beta_{\zeta \zeta S} (k_1,k_2,k_3)  \nonumber \\
	&&\hspace{20mm}  
	+ \{g_{T \ell_1}^{\rm (adi)}(k_1)g_{T \ell_2}^{\rm (iso)} (k_2) 
	g_{T \ell_3}^{\rm (iso)} (k_3) + \mbox{(cyclic with $\ell$'s)} \}\beta_{\zeta SS} (k_1,k_2,k_3)
	 \nonumber\\
	&&\hspace{20mm} \left. 
	+ g_{T \ell_1}^{\rm (iso)}(k_1)g_{T \ell_2}^{\rm (iso)} (k_2) 
	g_{T \ell_3}^{\rm (iso)} (k_3) \beta_{SSS} (k_1,k_2,k_3) \right] \nonumber \\
	&&\hspace{15mm} \times 
	\left[ P_{\delta \phi}(k_1)P_{\delta \phi}(k_2) + (2~{\rm perms})  \right], \label{b1}
\end{eqnarray}
where $j_\ell (x)$ is the spherical Bessel function.
Here we have assumed that the final terms in Eqs.~(\ref{beta1})-(\ref{beta4})
dominate. Otherwise, the above expression becomes more complicated.

As we found in the general curvaton model (see Sec.~\ref{gcm}),
requiring the isocurvature constraints $P_S/P_\zeta \lesssim 0.01$ for totally anti-correlated 
case,
the dominant contributions to the bispectrum come from the first two terms in Eq.~(\ref{b1}).
The first term, which is denoted by $b_{\ell_1 \ell_2 \ell_3}^{\rm (adi)}$, can be written as
\begin{eqnarray}
	&&b_{\ell_1 \ell_2 \ell_3}^{\rm (adi)}=
	\frac{8}{\pi^3}\int_0^\infty dr~r^2 \int_0^\infty dk_1~k_1^2
	g_{T \ell_1}^{\rm (adi)}(k_1)j_{\ell_1}(k_1r) P_{\delta \phi}(k_1)
	\int_0^\infty dk_2~k_2^2g_{T \ell_2}^{\rm (adi)}(k_2)j_{\ell_2}(k_2r) P_{\delta \phi}(k_2)
	\nonumber \\
	&&\hspace{15mm}\times \int_0^\infty dk_3~k_3^2 
	g_{T \ell_3}^{\rm (adi)} (k_3)j_{\ell_3}(k_3r)~
	\beta_{\zeta \zeta \zeta} +(2~{\rm perms})
	\nonumber \\
	&&\hspace{10mm}=\frac{48}{5\pi^3}\int_0^\infty dr~r^2 \int_0^\infty dk_1~k_1^2
	g_{T \ell_1}^{\rm (adi)}(k_1)j_{\ell_1}(k_1r) P_{\zeta}(k_1)
	\int_0^\infty dk_2~k_2^2g_{T \ell_2}^{\rm (adi)}(k_2)j_{\ell_2}(k_2r) P_{\zeta}(k_2)
	\nonumber \\
	&&\hspace{15mm} \times \int_0^\infty dk_3~k_3^2 
	g_{T \ell_3}^{\rm (adi)} (k_3)j_{\ell_3}(k_3r)~
	f_{\rm NL}^{(\rm adi)} +(2~{\rm perms}).	
	 \label{tem6}
\end{eqnarray}
The second term, which is denoted by $b_{\ell_1 \ell_2 \ell_3}^{\rm (cor1)}$, can be written as
\begin{eqnarray}
	&&b_{\ell_1 \ell_2 \ell_3}^{\rm (cor1)}= \left[ 
	\frac{8}{\pi^3}\int_0^\infty dr~r^2 \int_0^\infty dk_1~k_1^2
	g_{T \ell_1}^{\rm (adi)}(k_1)j_{\ell_1}(k_1r) P_{\delta \phi}(k_1)
	\int_0^\infty dk_2~k_2^2g_{T \ell_2}^{\rm (adi)}(k_2)j_{\ell_2}(k_2r) P_{\delta \phi}(k_2) \right.
	\nonumber \\
	&&\hspace{15mm} \left.\times \int_0^\infty dk_3~k_3^2 
	g_{T \ell_3}^{\rm (iso)} (k_3)j_{\ell_3}(k_3r)~
	\beta_{\zeta \zeta S} +(2~{\rm perms}) \right]+\mbox{(cyclic with $\ell$'s)}
	\nonumber \\ 
	&&\hspace{10mm}=\left[ \frac{144}{5\pi^3}\int_0^\infty dr~r^2 \int_0^\infty dk_1~k_1^2
	g_{T \ell_1}^{\rm (adi)}(k_1)j_{\ell_1}(k_1r) P_{\zeta}(k_1)
	\int_0^\infty dk_2~k_2^2g_{T \ell_2}^{\rm (adi)}(k_2)j_{\ell_2}(k_2r) P_{\zeta}(k_2) \right.
	\nonumber \\
	&&\hspace{15mm} \left. \times \int_0^\infty dk_3~k_3^2 
	g_{T \ell_3}^{\rm (iso)} (k_3)j_{\ell_3}(k_3r)~
	f_{\rm NL}^{(\rm cor1)} +(2~{\rm perms})\right]+\mbox{(cyclic with $\ell$'s)}.	
	 \label{tem6}
\end{eqnarray}
In deriving these equations, we have dropped logarithmic dependence on $k_1,k_2,k_3$
of $\beta_{\zeta \zeta \zeta}$ and $\beta_{\zeta \zeta S}$.
The third and forth terms, $b_{\ell_1 \ell_2 \ell_3}^{\rm (cor2)}$ and 
$b_{\ell_1\ell_2\ell_3}^{\rm (iso)}$, can also be written in the same way as
\begin{eqnarray}
	&&b_{\ell_1 \ell_2 \ell_3}^{\rm (cor2)}= \left[ 
	\frac{8}{\pi^3}\int_0^\infty dr~r^2 \int_0^\infty dk_1~k_1^2
	g_{T \ell_1}^{\rm (adi)}(k_1)j_{\ell_1}(k_1r) P_{\delta \phi}(k_1)
	\int_0^\infty dk_2~k_2^2g_{T \ell_2}^{\rm (iso)}(k_2)j_{\ell_2}(k_2r) P_{\delta \phi}(k_2) \right.
	\nonumber \\
	&&\hspace{15mm} \left.\times \int_0^\infty dk_3~k_3^2 
	g_{T \ell_3}^{\rm (iso)} (k_3)j_{\ell_3}(k_3r)~
	\beta_{\zeta S S} +(2~{\rm perms}) \right]+\mbox{(cyclic with $\ell$'s)}
	\nonumber \\ 
	&&\hspace{10mm}=\left[ \frac{432}{5\pi^3}\int_0^\infty dr~r^2 \int_0^\infty dk_1~k_1^2
	g_{T \ell_1}^{\rm (adi)}(k_1)j_{\ell_1}(k_1r) P_{\zeta}(k_1)
	\int_0^\infty dk_2~k_2^2g_{T \ell_2}^{\rm (iso)}(k_2)j_{\ell_2}(k_2r) P_{\zeta}(k_2) \right.
	\nonumber \\
	&&\hspace{15mm} \left. \times \int_0^\infty dk_3~k_3^2 
	g_{T \ell_3}^{\rm (iso)} (k_3)j_{\ell_3}(k_3r)~
	f_{\rm NL}^{(\rm cor2)} +(2~{\rm perms})\right]+\mbox{(cyclic with $\ell$'s)},
	 \label{tem7} \\
	&&b_{\ell_1 \ell_2 \ell_3}^{\rm (iso)}= 
	\frac{8}{\pi^3}\int_0^\infty dr~r^2 \int_0^\infty dk_1~k_1^2
	g_{T \ell_1}^{\rm (iso)}(k_1)j_{\ell_1}(k_1r) P_{\delta \phi}(k_1)
	\int_0^\infty dk_2~k_2^2g_{T \ell_2}^{\rm (iso)}(k_2)j_{\ell_2}(k_2r) P_{\delta \phi}(k_2) 
	\nonumber \\
	&&\hspace{15mm} \times \int_0^\infty dk_3~k_3^2 
	g_{T \ell_3}^{\rm (iso)} (k_3)j_{\ell_3}(k_3r)~
	\beta_{SSS} +(2~{\rm perms}) 
	\nonumber \\ 
	&&\hspace{10mm}= \frac{1296}{5\pi^3}\int_0^\infty dr~r^2 \int_0^\infty dk_1~k_1^2
	g_{T \ell_1}^{\rm (iso)}(k_1)j_{\ell_1}(k_1r) P_{\zeta}(k_1)
	\int_0^\infty dk_2~k_2^2g_{T \ell_2}^{\rm (iso)}(k_2)j_{\ell_2}(k_2r) P_{\zeta}(k_2) 
	\nonumber \\
	&&\hspace{15mm}  \times \int_0^\infty dk_3~k_3^2 
	g_{T \ell_3}^{\rm (iso)} (k_3)j_{\ell_3}(k_3r)~
	f_{\rm NL}^{(\rm iso)} +(2~{\rm perms}).
	 \label{tem8}
\end{eqnarray}

\begin{figure}[t]
 \begin{center}
   \includegraphics[width=1.0\linewidth]{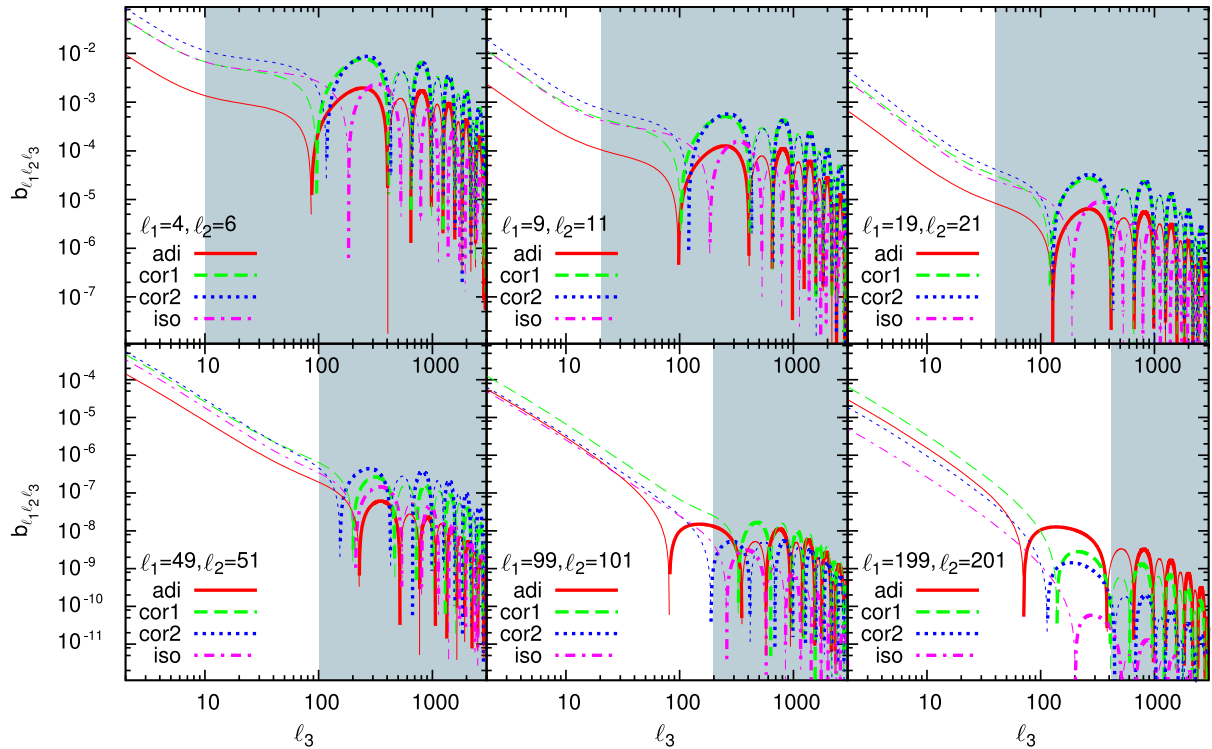}
   \caption{CMB bispectra $b^{\rm (adi)}_{\ell_1\ell_2\ell_3}$ (solid red line), 
   $b^{\rm (cor1)}_{\ell_1\ell_2\ell_3}$ (dashed green line)
   $b^{\rm (cor2)}_{\ell_1\ell_2\ell_3}$ (dotted blue line) and 
   $b^{\rm (iso)}_{\ell_1\ell_2\ell_3}$ (dot-dashed magenta line).
   The thick (thin) lines correspond to positive (negative) values of bispectra.
   We plotted $b_{\ell_1\ell_2\ell_3}$ as a function of $\ell_3$ with fixing 
   $(\ell_1,\ell_2)=(4,6), (9,11), (19,21), (49,51), (99,101), (199,201)$.   
   Unobservable multipoles are shown in the shaded regions.
   }
   \label{fig:bispec1}
 \end{center}
\end{figure}

\begin{figure}[t]
 \begin{center}
   \includegraphics[width=1.0\linewidth]{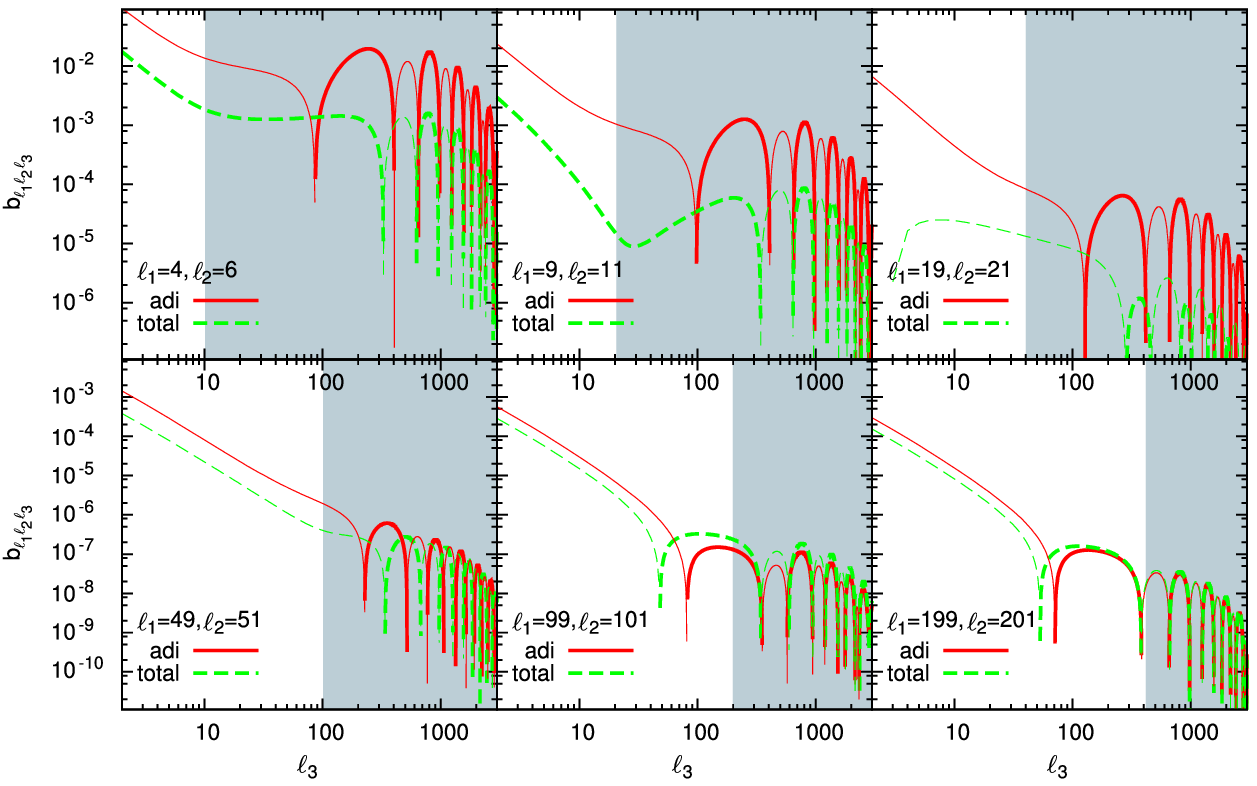}
   \caption{CMB bispectra $b_{\ell_1\ell_2\ell_3}$ (dashed green line) with 
   $f^{\rm (adi)}_{\rm NL} = 10, f^{\rm (cor1)}_{\rm NL} = -0.9, 
   f^{\rm (cor2)}_{\rm NL} = 0.03, f^{\rm (iso)}_{\rm NL} = -3\times10^{-4}$.
   For reference, we also plotted $b^{\rm (adi)}_{\ell_1\ell_2\ell_3}$ (solid red line).
   Same as in Fig.~\ref{fig:bispec1}, 
   the thick (thin) lines correspond to positive (negative) values of bispectra and 
   unobservable multipoles are shown in the shaded regions.
   }
   \label{fig:bispec2}
 \end{center}
\end{figure}

We now move to see how non-Gaussianity from correlated isocurvature
and adiabatic perturbations shows its signature on CMB bispectra $b_{\ell_1\ell_2\ell_3}$.
To see this we plotted CMB bispectra $b_{\ell_1 \ell_2 \ell_3}$ 
in Fig.~\ref{fig:bispec1} and \ref{fig:bispec2}
using a modified version of the publicly available CMB code {\tt camb} \cite{Lewis:1999bs}. 
Here we assumed 
a flat scale-invariant SCDM model ($\Omega_{\rm m} = 1$) and 
adopted a following set of cosmological parameters 
$(\Omega_{\rm b}=0.05, \Omega_{\rm cdm}=0.95, H_0 = 50)$, where
$\Omega_{\rm b}$ and $\Omega_{\rm cdm}$ are the energy density of baryon and CDM, and 
$H_0$ is the Hubble parameter in unit of km/s/Mpc.
For simplicity in visualization, 
we have chosen following sets of $(\ell_1,\ell_2) = (4,6), (9,11), (19,21), 
(49,51), (99,101), (199, 201)$ and shown $b_{\ell_1\ell_2\ell_3}$ as a function of $\ell_3$. 
Fig.~\ref{fig:bispec1} shows bispectra separately with of their dependences on initial perturbations, 
$b^{\rm (adi)}_{\ell_1\ell_2\ell_3},~b^{\rm (cor1)}_{\ell_1\ell_2\ell_3},~
b^{\rm (cor2)}_{\ell_1\ell_2\ell_3},~b^{\rm (iso)}_{\ell_1\ell_2\ell_3}$, where
we fixed $\beta_{\zeta\zeta\zeta}=\beta_{\zeta\zeta S}=
\beta_{\zeta SS}=\beta_{SSS}=1$.
We can see from Fig.~\ref{fig:bispec1} that the bispectra with isocurvature, 
$b^{\rm (cor1)}_{\ell_1\ell_2\ell_3},~
b^{\rm (cor2)}_{\ell_1\ell_2\ell_3}$ and $b^{\rm (iso)}_{\ell_1\ell_2\ell_3}$, all show that
their amplitudes are larger at large angular scales than that of the adiabatic bispectrum
$b^{\rm (adi)}_{\ell_1\ell_2\ell_3}$.
This can be easily understood that the transfer function for isocurvature  perturbation
$g^{\rm (iso)}_\ell(k)$ is large at large angular scales ($\ell \lesssim 10$) than 
adiabatic one $g^{\rm (adi)}_\ell(k)$ (See \cite{Kawasaki:2008sn} for more detailed discussions).
At small scales isocurvature perturbation tends to give small amplitude on CMB anisotropies. 
However, in some specfic configurations, bispectra arising from correlations of adiabatic 
and isocurvature give comparable or even larger amplitudes 
than the pure adiabatic bispectrum at relatively small scales.
Especially, $b^{\rm (cor1)}_{\ell_1\ell_2\ell_3}$
is large when two of $\ell$'s are in regime of acoustic oscillation (e.g. $\ell_1,\ell_2\simeq 200$) 
and the other $\ell$ is small ($\ell_3 \lesssim 10$), where 
$g^{\rm (adi),~(iso)}_\ell(k)$ of their largest amplitudes are picked out. 

In Fig.~\ref{fig:bispec2} we plotted total bispectra $b_{\ell_1\ell_2\ell_3}$
with realistic values of $f^{\rm (adi)}_{\rm NL}=10, f^{\rm (cor1)}_{\rm NL}=-0.9,
f^{\rm (cor2)}_{\rm NL}=0.03, f^{\rm (iso)}_{\rm NL}=-3\times10^{-4}$, which, for example, 
can be realized
by taking $p=0, R=0.1, \epsilon_X=0.03$ in Eqs.~(\ref{eq:fNL1}-\ref{eq:fNL4}).
We can see there are considerable differences from pure adiabatic bispectra,
which we have plotted in Fig.~\ref{fig:bispec2} for reference.
The most dominant contribution for deviating from pure adiabatic bispectra
comes from the bispectra from the correlation of two adiabatic and 
one isocurvature inital perturbations, $b^{\rm (cor1)}_{\ell_1\ell_2\ell_3}$ since 
$f^{\rm (cor1)}_{\rm NL}$ 
is at least order of magnitude larger than $f^{\rm (cor2)}_{\rm NL}$ and $f^{\rm (iso)}_{\rm NL}$ 
so as not to conflict with current constraints on isocurvature perturbations. 
Therefore we can conclude that the signature of non-Gaussianity from correlated isocurvature
and adiabatic perturbations are found in CMB bispectra at least one of $\ell$'s is 
small, where isocurvature perturbation gives large amplitude on CMB anisotropies. 
Also at intermediate angular scales ($\ell\simeq 100$), where 
bispectra shows acoustic oscillation but still have sufficient power from isocurvature perturbation, 
we can see the oscillation of $b_{\ell_1\ell_2\ell_3}$ has different phase from that of  pure adiabatic
one $b^{\rm (adi)}_{\ell_1\ell_2\ell_3}$, which would be a striking evidence for correlated isocurvature
non-Gaussianity.


\section{Discussion and conclusions} \label{sec:conclusion}

In this paper we have generalized our formalism provided in Ref.~\cite{Kawasaki:2008sn}
for calculating non-Gaussianity,
to include more general case where correlation between adiabatic and isocurvature perturbations exist,
and shown that it can significantly affect the bispectrum of the CMB anisotropy.
Actually in the curvaton scenario, the correlated isocurvature perturbation between 
CDM and radiation exists, unless all the dark matter arise after the curvaton decays.
Although we have focused on the CDM isocurvature perturbation,
similar mechanism can produce the baryonic isocurvature perturbation, if the 
baryon number is created before the curvaton decays.
Therefore, the standard prediction for $f_{\rm NL}$ in the curvaton/ungaussiton scenario correctly 
characterizes the non-Gaussian properties only when no isocurvature perturbations exist.
In other words, there is a chance to probe the physics in the early Universe by using
non-Gaussian isocurvature perturbations, if detected.

The formulations provided in this paper can be applied to broad class of models.
The right-handed sneutrino ($\tilde N$) may be light during inflation and 
acquire quantum fluctuations.
The decay of $\tilde N$ generates lepton number, as well as some fraction of the total radiation. 
If the fluctuation of $\tilde N$ significantly contributes to the curvature perturbation,
it leaves the correlation between the adiabatic and baryonic isocurvature perturbations.
The AD field can have similar effect.
The modulated reheating scenario \cite{Dvali:2003em} also predicts mixture of adiabatic and
isocurvature perturbations with correlations if the CDM/baryon is created
before the inflaton decays.

It is obvious that non-Gaussianity in the cosmological perturbations provides invaluable information
on the very early Universe.
It would be interesting if the future detection of the non-Gaussainity tells us about the origin of
CDM and/or baryon asymmetry of the Universe, through their small isocurvature perturbations.
\\

{\it Note added}\\
While finalizing this manuscript, Ref.~\cite{Langlois:2008vk} appeared in the preprint server,
which treats similar subjects to ours.

\section*{Acknowledgment}

K.~Nakayama and T.~Sekiguchi would like to thank the Japan Society for the Promotion of
Science for financial support.
This work is supported by Grant-in-Aid for Scientific research from the Ministry of Education,
Science, Sports, and Culture (MEXT), Japan, No.14102004 (M.K.)
and also by World Premier International
Research Center InitiativeiWPI Initiative), MEXT, Japan.



{}

\end{document}